\documentclass[twocolumn,preprintnumbers,amsmath,amssymb]{revtex4}
\usepackage{graphicx}
\usepackage{dcolumn}
\usepackage{bm}
\newcommand{\comment}[1]{}
\newcommand\etal{\mbox{\textit{et al.}}}

\begin{document}
\setlength{\unitlength}{0.7\textwidth}
\preprint{}

\title{Inertial range scaling of scalar flux spectra in uniformly sheared turbulence}


\author{W.J.T. Bos (correspondent)}
\author{J.-P. Bertoglio}

\affiliation{%
 LMFA, UMR CNRS 5509\\ Ecole Centrale de Lyon - Universit\'e Lyon 1 - INSA Lyon\\ Ecully, France}

\begin{abstract}
A model based on two-point closure theory of turbulence is proposed and applied to study the Reynolds number dependency of the scalar flux spectra in homogeneous shear flow with a cross-stream uniform scalar gradient. For the cross-stream scalar flux, in the inertial range the spectral behavior agrees with classical predictions and measurements. The streamwise scalar flux is found to be in good agreement with the results of atmospherical measurements. However, both the model results and the atmospherical measurements disagree with classical predictions. A detailed analysis of the different terms in the evolution equation for the streamwise scalar flux spectrum shows that nonlinear contributions are governing the inertial subrange of this spectrum and that these contributions are relatively more important than for the cross-stream flux. A new expression for the scalar flux spectra is proposed. It allows to unify the description of the components in one single expression, leading to a classical $K^{-7/3}$ inertial range range for the cross-stream component and to a new $K^{-23/9}$ scaling for the streamwise component that agrees better with atmospherical measurements than the $K^{-3}$ prediction of Wyngaard and Cot\'e [Quart. J. R. Met. Soc., {\bf 98}, 590 (1972)].
\end{abstract}

\maketitle

\section{Introduction}

Homogeneous turbulence subjected to uniform shear is one of the classical problems in turbulence. The understanding and modeling of mixing of a passive scalar in such flows is of fundamental interest for the prediction of atmospherical flows, dispersion problems and heat transfer. A key quantity in the mixing and transport process is the turbulent scalar flux $\overline{u_i\theta}$, with $u_i$ the velocity fluctuation and  $\theta$ the scalar fluctuation. It is the quantity that is accounting for the effects of turbulence on the mean scalar profile. In order to characterize the contributions of the different turbulent lengthscales to the scalar flux, it is important to analyze the spectral distribution of scalar flux over wavenumbers, i.e. the scalar flux spectrum. In the literature, the spectral distribution of scalar flux produced by a uniform scalar gradient imposed on isotropic turbulence, was extensively studied.\cite{Lumley,Mydlarski,Mydlarski2,GormanJFM,Bos2005}  In the present paper attention is focused on the case of turbulence in the presence of a constant shear $S$ and a cross-stream uniform scalar gradient $\Gamma$. The scalar and velocity gradients are chosen in the same direction, that we call the $z$-axis. The mean flow is in the $x$ direction: 
\begin{equation}\label{grads}
S=\frac{\partial \overline{U}_x}{\partial z},~~~\Gamma=\frac{\partial {\overline{\Theta}}}{\partial z}
\end{equation}
with $\overline{\Theta}$ the mean scalar. The  only non-zero components of the scalar flux are in this case $\overline{w\theta}$ and $\overline{u\theta}$, the cross-stream and streamwise scalar fluxes respectively. The three-dimensional spectra corresponding to these quantities are defined by
\begin{eqnarray}
\mathcal{F}_{u_i\theta}(\bm K)={FT_{/\bm r}\left[\overline{u_i(\bm{x})\theta(\bm{x+r})}\right]}\textrm{,}
\end{eqnarray}
in which $\bm K$ is the wavevector and $FT_{/\bm r}$ denotes a Fourier transform with respect to the separation vector $\bm r$. In the present paper attention is focused on the spherically averaged spectra:
\begin{eqnarray}\label{eqf3}
F_{u_i\theta}(K)=\int_{\Sigma(K)}{\mathcal{F}_{u_i\theta}(\bm K)}d\Sigma(K)\textrm{,}
\end{eqnarray}
{
where $\int_{\Sigma(K)}~.~d\Sigma(K)$ stands for integral over a spherical shell with a radius $K$,}
the wavenumber. For large Reynolds numbers in the inertial range  the cross-stream scalar flux spectrum is generally believed to obey:
\begin{equation}\label{lumleyscaling}
F_{w\theta}(K)\sim \frac{\partial {\overline{\Theta}}}{\partial z}\epsilon^{1/3}K^{-7/3},
\end{equation}
 as was proposed by Lumley \cite{Lumley2}. $\epsilon$ is the spectral energy flux. This scaling is identical to the one for the unsheared case, in which isotropic turbulence is interacting with a scalar gradient (see Bos \etal \cite{Bos2005} for a discussion of this case).

For the streamwise scalar flux spectrum, Wyngaard and Cot\'e \cite{Wyngaard} proposed the expression:
\begin{equation}\label{WC-3}
F_{u\theta}(K)\sim \frac{\partial {\overline{\Theta}}}{\partial z} SK^{-3}.
\end{equation}
This scaling was however never clearly observed. The atmospheric measurements of Kaimal {\it et al.} \cite{Kaimal}, Kader and Yaglom \cite{Kader} and Caughey \cite{Caughey} of the streamwise heat flux spectrum show an inertial exponent closer to $-2.5$ than to the $-3$ resulting from the analysis by Wyngaard and Cot\'e \cite{Wyngaard}. This discrepancy between the classical scaling and observations motivates the present work. 

In section \ref{secModel} the two-point closure approach described in Bos \etal \cite{Bos2005} will be extended to the case of uniform shear flow. The results of the approach, based on the EDQNM (Eddy-Damped Quasi-Normal Markovian) theory are presented in section \ref{secResults}, and the behavior of the spectrum as the Reynolds number is increased is discussed. A detailed analysis of the spectra is performed and the scalings observed in the inertial range are shown to agree with experimental observations. In section \ref{secDima} a new expression is proposed for the streamwise scalar flux spectrum using dimensional arguments.   

\section{Modeling the scalar flux spectrum \label{secModel}}

Experimental difficulties limit the study of homogeneous shear flows both in Reynolds number $R_\lambda$ and non dimensional time $St$. Information at higher $R_\lambda$ can be obtained from atmospheric measurements in which the Reynolds number is typically within the range $10^3<R_\lambda<10^4$ (see for example the paper of Bradley {\it et al.} \cite{Bradley}). Atmospheric flows, although generally subject to shear, can however hardly be considered to be subject to a uniform shear and a major problem with atmospheric experiments is the uncontrolable nature of the meteorological conditions. The mean shear is generally not constant over a long interval of time, the initial and boundary conditions are not easily known by the experimentalist and the role played by buoyancy effects can also be difficult to assess.

Exactly the opposite is the case with direct numerical simulations (DNS): the initial conditions as well as the boundary conditions are entirely determined by the scientist or the engineer. The limits of these 'numerical experiments' are only dependent on the available computer resources. These computational limits restricted the DNS study of Rogers {\it et al.} \cite{RogersNASA} to  $R_\lambda\approx 40$, a Reynolds number at which an inertial range is not clearly observable. Higher resolution DNS were recently performed by Brethouwer \cite{Brethouwer2005}, but the results for the scalar flux spectra were not presented in this work.  The Reynolds number limitation can be removed by performing Large Eddy Simulations (LES) in which only the large scales are resolved: we mention the work of Kaltenbach {\it et al.} \cite{Kaltenbach}. The problems are then that the filtering operation and the subgrid model can affect the results in the inertial range. Another problem with both DNS and LES is that in homogeneous shear flow the integral lengthscale increases monotonically (and much faster than in the case of isotropic turbulence) so that, quite rapidly, its size will become equal to the one of the computational domain. 

In a recent paper \cite{Bos2005}, it was shown that the EDQNM approach is very suitable to study the Reynolds number dependency of the inertial range of the scalar flux spectrum. The study was developed for the case of isotropic turbulence. The approach is here extended to homogeneous shear flow. In the case of a fluctuating scalar field produced by the interaction of an isotropic turbulence with a uniform mean scalar gradient, there is an exact relation between the 3D spectrum $\mathcal{F}_{u_i\theta}(\bm K)$ and its integral over a sphere with radius $K$ (Herr \emph{et al.} \cite{Herr}):
\begin{equation}\label{eqFmu}
\mathcal{F}_{u_i\theta}(\bm K)\sim (1-\mu^2)F_{u_i\theta}(K)
\end{equation}
with $\mu$ the cosine of the angle between the scalar gradient axis and the wavevector. In the presence of shear, relation (\ref{eqFmu}) does not hold anymore; nor the spectral tensor $\Phi_{ij}$ can be expressed exactly as a function of the wavenumber only. A full EDQNM approach of the sheared problem would then require to build and numerically integrate a wavevector dependent closed set of equations\cite{Berto81}. In order to simplify the numerical task that would result of this complete approach, the equations are integrated over spherical shells with radius $K$ so that the basic quantities in the present approach will still be the spherically averaged spectra $F_{u_i\theta}(K)$. The difference with the unsheared case will be that the shell averaged description is now a simplification that will require more closure assumptions in the model derivation.

The main quantities in this paper are then the spherically averaged spectra: $F_{u_i\theta}(K)$ defined by (\ref{eqf3}) and $\varphi_{ij}(K)$ defined as:
\begin{eqnarray}\label{phiuiuj}
\varphi_{ij}(K)=\int_{\Sigma(K)}{\Phi_{ij}(\bm K)}d\Sigma(K)\textrm{,}
\end{eqnarray}
{
with}  
\begin{eqnarray}
\Phi_{ij}(\bm K)={FT_{/\bm r}\left[\overline{u_i(\bm{x})u_j(\bm{x+r})}\right]}.
\end{eqnarray}
The approach consisting in considering and modeling only the spherically averaged spectra was introduced by Cambon \emph{et al.} \cite{Cambon81} for the velocity field. The averaging procedure introduces unclosed terms in the equations of both $\varphi_{ij}(K)$ and $F_{u_i\theta}(K)$. The unclosed terms that are related to the interaction with the mean velocity field (linear transfer and rapid pressure terms) are modeled by tensor invariant theory. The nonlinear terms are modeled by the EDQNM theory with additional assumptions about the anisotropy of the spectra.

We will calibrate the resulting model by comparison with experimental results on the turbulent Prandtl number and apply it to flows with  Reynolds numbers ranging from low values, attainable by DNS and laboratory experiments, up to values that correspond to the highest Reynolds numbers observed in atmospheric measurements. We note that the 
{
molecular
}
Prandtl number in this paper is taken equal to unity.

\subsection{The equation for the scalar flux spectrum }

In homogeneously sheared turbulence in the presence of a uniform mean scalar gradient, the equation for the 3D scalar flux spectrum is 

\begin{eqnarray}\label{eqcrayaF}
\left[ \frac{\partial}{\partial t}+(\nu+\alpha) K^2\right]\mathcal{F}_{u_i\theta}
+\frac{\partial \overline{U}_i}{\partial x_j}\mathcal{F}_{u_j\theta}
+\frac{\partial \overline{\Theta}}{\partial x_j}\Phi_{ij}=\nonumber \\
+\frac{\partial \overline{U}_n}{\partial x_j} \frac{\partial K_n \mathcal{F}_{u_i\theta}}{\partial K_j}
+2\frac{\partial \overline{U}_n}{\partial x_j}                      
\frac{K_i K_n}{K^2}\mathcal{F}_{u_j\theta}\nonumber \\
-iK_n\left(T_{\theta in}-T_{\theta in}^* \right) + i\frac{K_iK_nK_j}{K^2}T_{\theta jn}
\end{eqnarray}
In this equation $\mathcal{F}_{u_j\theta}$, $\Phi_{ij}$ and the two-point triple correlations $T_{\theta in}$, 
{
\begin{eqnarray}
T_{\theta in}=FT_{/\bm r}\Big(\overline{\theta(\bm x)u_i(\bm x)u_n(\bm x+\bm r)}\Big),
\end{eqnarray}}
are functions of the wavevector and time. $\nu$ and $\alpha$ are the molecular viscosity and diffusivity respectively. By integration of equation (\ref{eqcrayaF}) over spherical shells of radius $K$, the equation for ${F}_{u_i\theta}$ is obtained  
\begin{eqnarray}\label{eqIntcrF}
\left[ \frac{\partial}{\partial t}+(\nu+\alpha) K^2\right]{F}_{u_i\theta}(K,t)
+\frac{\partial \overline{U}_i}{\partial x_j}{F}_{u_j\theta}(K,t)=\nonumber\\
-\frac{\partial \overline{\Theta}}{\partial x_j}\varphi_{ij}(K,t)
+T_{i\theta}^L(K,t)+\Pi_{i\theta}^L(K,t)\nonumber\\
+T_{i\theta}^{NL}(K,t)+\Pi^{NL}_{i\theta}(K,t).
\end{eqnarray}
The terms on the left hand side of this equation  do not require modeling: the dissipation and the production by shear are closed terms. On the RHS we find a production term involving the spherically averaged spectral tensor $\varphi_{ij}(K,t)$ defined in expression (\ref{phiuiuj}), and two terms involving explicitly the mean shear. The first term in which the mean shear appears is the linear transfer 
{
\begin{eqnarray} \label{LinTran0}
T_{i\theta}^L=\frac{\partial \overline{U}_n}{\partial x_j }\int_{\Sigma(K)} \frac{\partial K_n \mathcal{F}_{u_i\theta}}{\partial K_j} d\Sigma(K).
\end{eqnarray}
This term is directly related to the stretching of material elements by the mean velocity gradient. It corresponds to a pure transfer in wavenumber space, as its integral over $K$ is zero. The second term involving the mean shear is the rapid pressure term 
\begin{eqnarray} \label{rapid0}
\Pi_{i\theta}^L=2\frac{\partial \overline{U}_n}{\partial x_j }\int_{\Sigma(K)}\frac{K_i K_n}{K^2}\mathcal{F}_{u_j\theta} d\Sigma(K).
\end{eqnarray}
It corresponds to the linear part of the pressure-scalar fluctuation correlation. The nonlinear transfer and nonlinear pressure terms are respectively defined as
\begin{eqnarray}
T_{i\theta}^{NL}(K,t)=\int_{\Sigma(K)}iK_n\left(T_{\theta in}^*-T_{\theta in} \right)d\Sigma(K),\\
\Pi_{i\theta}^{NL}(K,t)=\int_{\Sigma(K)}i\frac{K_iK_nK_j}{K^2}T_{\theta jn}d\Sigma(K).
\end{eqnarray}
The integral of $T_{i\theta}^{NL}(K,t)$ over all wavenumbers is zero and this term corresponds therefore to a flux in spectral space (cascade term). The nonlinear pressure term is a destructive term that tends to decorrelate the scalar and velocity fluctuations. 
}
{
These two terms could be combined as they both arise from the nonlinear triple correlation $T_{\theta ij}$. It is however informative to separate the two contributions, as it allows a more precise analysis of the results, as will be shown in section \ref{Contri}.
}
All the terms on the RHS of (\ref{eqIntcrF}) are to be modeled. In the following the dependence on $K$ and $t$ will be omitted from the equations.

\subsection{Production by the mean scalar gradient}

\begin{figure}
\includegraphics[scale=0.65,angle=0]{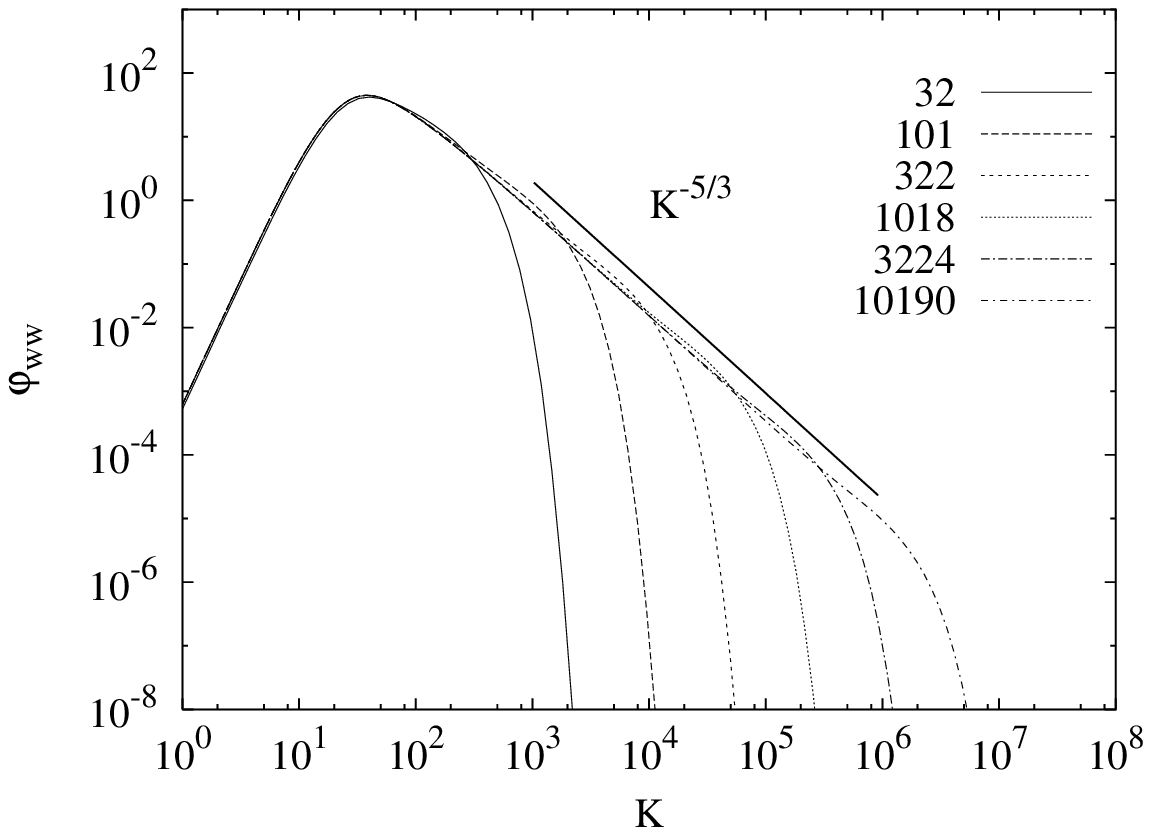}
\includegraphics[scale=0.65,angle=0]{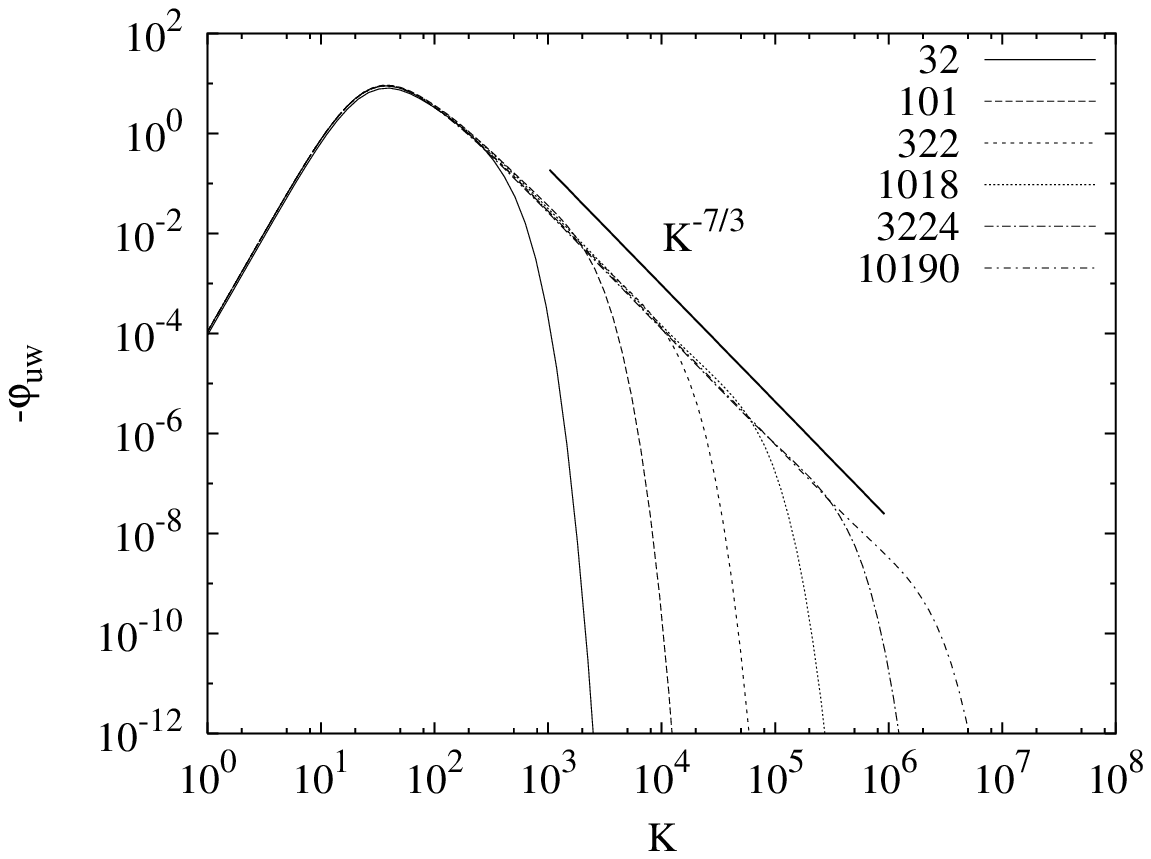}
\caption{Velocity spectra $\varphi_{ww}(K)$ and $\varphi_{uw}(K)$ for $32<R_{\lambda}<10^4$ at $St=0.5$. \label{specE33E13}}
\end{figure}

In equation (\ref{eqIntcrF}) the production by the mean scalar gradient involves the spherically averaged spectral tensor $\varphi_{ij}$. Choosing the orientation as given in expression (\ref{grads}) the only relevant components are $\varphi_{ww}$ and $\varphi_{uw}$. The spherically averaged spectral tensor is modeled by an extension of the EDQNM theory for anisotropic turbulence. 
{
The extension of this type of closure to anisotropic turbulence using spherically averaged quantities was first proposed by Cambon \emph{et al.} \cite{Cambon81}. In the present approach, as in Cambon \emph{et al.} the equation for $\varphi_{ij}$ is solved. The unknown terms appearing in this equation are modeled as
proposed by Cambon \emph{et al.} for the pressure contribution, and using simpler expressions for the transfer terms (Touil \emph{et al.}   \cite{Touil3}).
}
 A detailed discussion of the results of this model is beyond the scope of this paper. In the present work we  only show the relevant spectra ($\varphi_{ww}$ and $\varphi_{uw}$) for varying Reynolds number. These are presented in figure \ref{specE33E13} for $St=0.5$. It can be observed that, as the Reynolds number increases, the slopes of the spectra in the inertial range are approaching the classical values of $-5/3$ and $-7/3$ for the diagonal and non diagonal components respectively.   

\subsection{Rapid pressure}

The rapid pressure term appearing in the scalar flux spectrum equation (eq. \ref{eqIntcrF}) is:
\begin{eqnarray} \label{rapid}
\Pi_{i\theta}^L=2\frac{\partial \overline{U}_n}{\partial x_j }\int_{\Sigma(K)}\frac{K_i K_n}{K^2}\mathcal{F}_{u_j\theta} d\Sigma(K)
\end{eqnarray}
At the level of description of the present model, the integral of $\mathcal{F}_{u_j\theta}$ is known, but not the integral of its moment:
\begin{eqnarray}\label{Hijn}
H^j_{in}=\frac{K_i K_n}{K^2}\mathcal{F}_{u_j\theta}.
\end{eqnarray}
In order to express $\Pi_{i\theta}^L$, the tensor $H^j_{in}$ is modeled assuming that it can be represented  as an isotropic tensorial function of $F_{u_i\theta}$ and $\alpha_i=K_i/K$. This does not imply that $H^j_{in}$, or the resulting model for the rapid pressure, are isotropic, but only that their anisotropic character reflects the anisotropy of the $F_{u_i\theta}$ vector. Using the theory of invariants and the representation by tensorial isotropic functions (as in Eringen \cite{Eringen} or Schiestel \cite{Schiestel}) and retaining only the lowest order terms, it is found
\begin{eqnarray}\label{eqH}
 H^j_{in}=A^L\delta_{in}{F}_{u_j\theta}+B^I\delta_{ij}F_{u_n\theta} +B^{II}\delta_{jn}F_{u_i\theta}\nonumber\\
+C\alpha_i\alpha_n{F}_{u_j\theta}+D^I\alpha_i\alpha_jF_{u_n\theta}   +D^{II}\alpha_j\alpha_nF_{u_i\theta} 
\end{eqnarray}
This expression needs to satisfy symmetry of the indices $i,n$, incompressibility and the definition of  ${F}_{u_j\theta}$, expression (\ref{eqf3}). These constraints allow to reduce the number of constants from 6 to 1.
The resulting expression for the rapid pressure term is
\begin{eqnarray}\label{rappressF}
\Pi_{i\theta}^L=(12A^L-4)\frac{\partial \overline{U}_i}{\partial x_j }{F}_{u_j\theta}-(8A^L-3)\frac{\partial \overline{U}_j}{\partial x_i }{F}_{u_j\theta}.
\end{eqnarray}

\subsection{Linear transfer \label{sectlinF}}

The problem of modeling the linear transfer is strongly connected to the modeling of the rapid pressure term. As a matter of fact, after some algebra, it is straightforward to show that it only requires a modeled expression for $H^j_{in}$, a problem that has been solved in the previous section.

The linear transfer is:
\begin{eqnarray} \label{LinTran}
T_{i\theta}^L=\frac{\partial \overline{U}_n}{\partial x_j }\int_{\Sigma(K)} \frac{\partial K_n \mathcal{F}_{u_i\theta}}{\partial K_j} d\Sigma(K)
\end{eqnarray}
This expression can be rewritten as:
\begin{eqnarray} \label{LinTran2}
T_{i\theta}^L=\frac{\partial \overline{U}_n}{\partial x_j }\frac{\partial}{\partial K}\int_0^K\int_{\Sigma(K)} \frac{\partial K_n \mathcal{F}_{u_i\theta}}{\partial K_j} d\Sigma(K)dK.
\end{eqnarray}
Noticing that the double integral can be rewritten as an integral over the volume of a sphere with radius $K$ and  applying the Gauss divergence theorem to this sphere, expression (\ref{LinTran2}) can be rewritten. The result has the form of a surface integral involving the outer vector $K_j/K$:
\begin{eqnarray} \label{LinTran4}
T_{i\theta}^L=\frac{\partial \overline{U}_n}{\partial x_j }\frac{\partial}{\partial K} K\int_{\Sigma(K)} H^i_{jn}  d\Sigma(K)
\end{eqnarray}
with $H^i_{jn}$ defined by (\ref{Hijn}). Using the modeled expression  (\ref{eqH}), the closed form for the linear transfer reads:
\begin{eqnarray}\label{eqlintransF}
T_{i\theta}^L=\left(\frac{3}{2}-4A^L\right)\left(\frac{\partial \overline{U}_i}{\partial x_j }+\frac{\partial \overline{U}_j}{\partial x_i}\right)\frac{\partial }{\partial K}K {F}_{u_j\theta}(K).
\end{eqnarray}
It has to be noticed that the approach used in this paper to model the linear transfer is similar to the one proposed by Cambon {\it et al.} \cite{Cambon81,Cambon} (see also Clark and Zemach \cite{Clark}) for the linear transfer of kinetic energy.
{
It can be argued that even if expression  (\ref{eqH}) may not necessarily be a good model for the tensor $H^i_{jn}$ itself, it could still lead to a satisfactory representation of the spherically integrated expressions (\ref{rapid}) and (\ref{LinTran}). It is also stressed that the exact forms of the rapid pressure and linear tranfer terms do not influence the asymptotic scalings of the spectra in the inertial range that are essentially determined by the nonlinear interactions.}

\subsection{Nonlinear transfer and nonlinear pressure}

In the presence of shear, the nonlinear transfer and pressure term can not be expressed exactly as a function of the wavenumber only. A full EDQNM approach of the problem would then require to build and numerically integrate a wavevector dependent closed set of equations. This approach is complex and numerically expensive. We will here treat the nonlinear transfer and slow pressure terms with the EDQNM model, derived for unsheared turbulence, using the formulation of  Bos \etal \cite{Bos2005} Applying this model, originally developed by Herr \etal \cite{Herr} for isotropic turbulence, to the case of an anisotropic velocity field has to be considered as an approximation. 
{
By introducing this approximation, a detailed description of anisotropy in wavevector-space is lost. The approach is justified in the limit of a small shear and is here used for strong shear as an extrapolation.   
}
The expressions for the nonlinear terms $T^{NL}$ and $\Pi^{NL}$ are not reproduced here (Equations (14) and (15) of reference \onlinecite{Bos2005}). These closed terms are exactly the same as in reference \onlinecite{Bos2005} introducing two model constants $\lambda'$ and $\lambda''$.

\section{Results \label{secResults}}

\subsection{Calibration}

The model for the scalar flux spectrum contains three constants. Two of these constants are the ones involved in the EDQNM closure  as mentioned in the previous section. These constants were determined in reference \onlinecite{Bos2005} and are not changed in the present work ($\lambda'=0.52$ and $\lambda''=0$), to satisfy compatibility with reference \onlinecite{Bos2005} in the case of vanishing shear. The third constant is $A^L$, the one that appears in the linear transfer and rapid pressure (expression (\ref{rappressF}) and (\ref{eqlintransF})). This last constant is determined by comparison of the value of the turbulent Prandtl number with the wind tunnel  experiments of Tavoularis and Corrsin \cite{Tavoularis1}. 
{
In this experiment the mean shear is generated by using an array of parallel channels of different mean flow with cylindrical heating rods at their exits. 
}
The shear rate in the simulations  is chosen equal to the one in the experiment, $S=47 s^{-1}$. The turbulent velocity field, assumed to be initially isotropic, is initialized by a {\it von K\'arm\'an} spectrum.\cite{Hinze} This initial spectrum is chosen so that the Reynolds number at $St=12$ is $R_\lambda\approx 150$, corresponding to the experimental values ($130<R_\lambda<160$). The scalar fluctuation is initially zero. 
{
Nonzero scalar fluctuations would not influence the scalar fluxes as long as they are initially isotropically distributed. 
}

The turbulent Prandtl number is the ratio of the Reynolds stress to the cross-stream  turbulent scalar flux, normalized by the mean gradients:
\begin{eqnarray}
Pr_T=\frac{\Gamma}{S}\frac{\overline{uw}}{\overline{w\theta}}.
\end{eqnarray}
Best agreement for this quantity with the Tavoularis and Corrsin \cite{Tavoularis1} experiment ($Pr_T=1.1$ at $St=12$) is obtained for a value of $A^L=0.385$. This value is retained in the following. We note that the spectral exponents are not sensitive to the choice of this constant.

\subsection{Spectra and Reynolds number dependency\label{secSpecresults}}

\begin{figure}
\includegraphics[scale=0.65,angle=0]{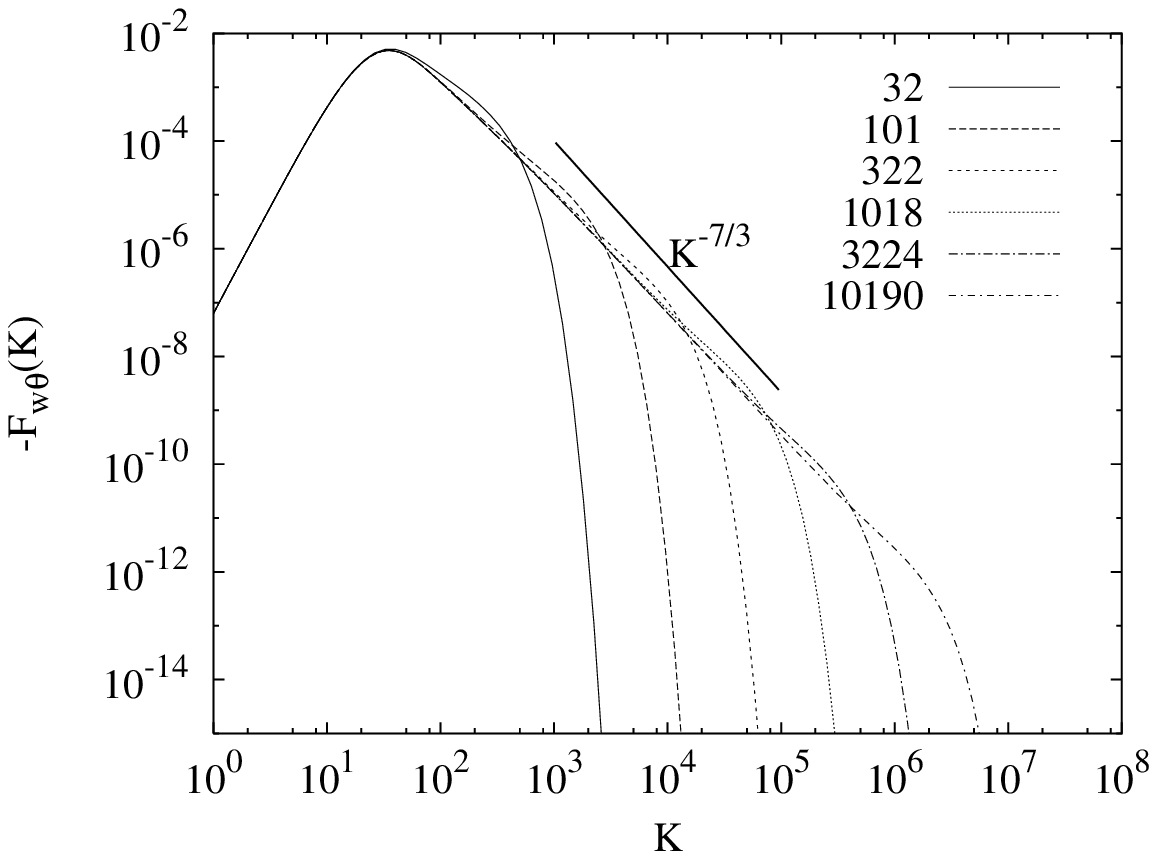}
\includegraphics[scale=0.65,angle=0]{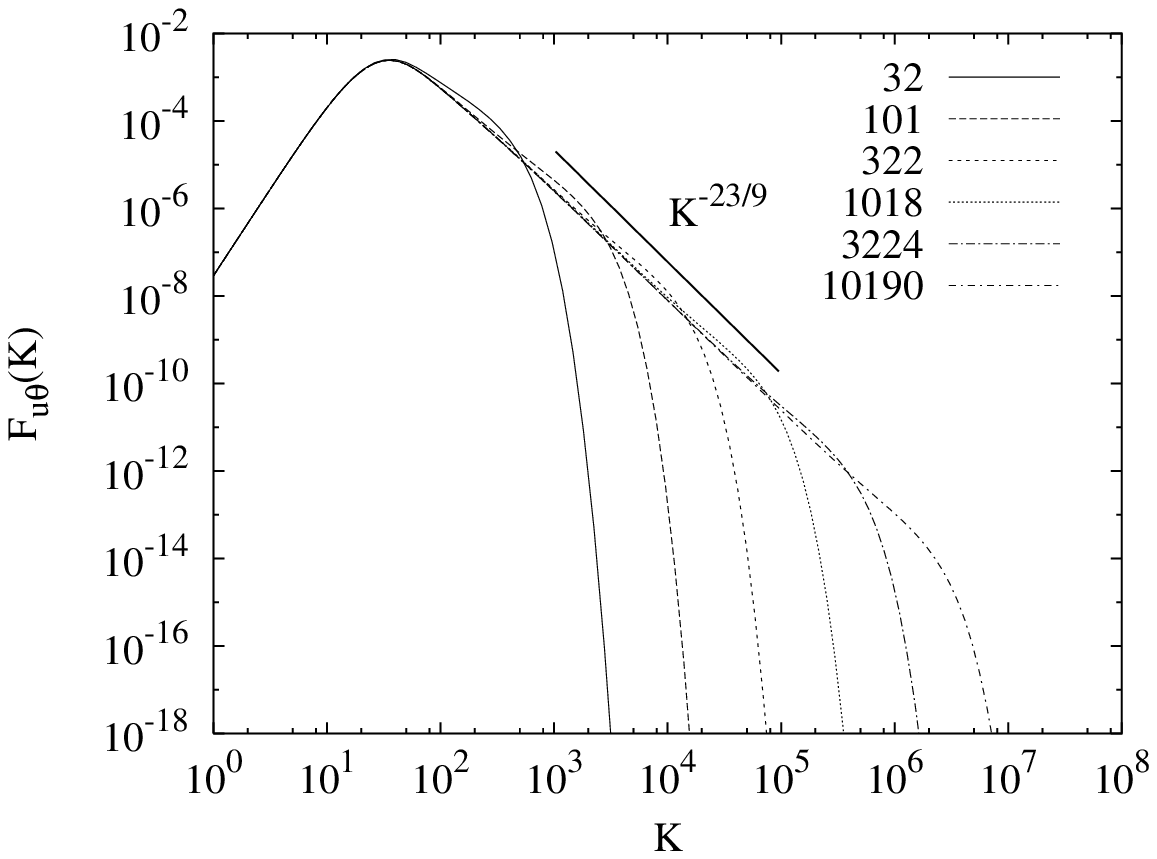}
\caption{Cross-stream and streamwise scalar flux spectra at $St=0.5$ for $32<R_{\lambda}<10^4$. \label{specF1F2} }
\end{figure}

\begin{figure}
\includegraphics[scale=0.65,angle=0]{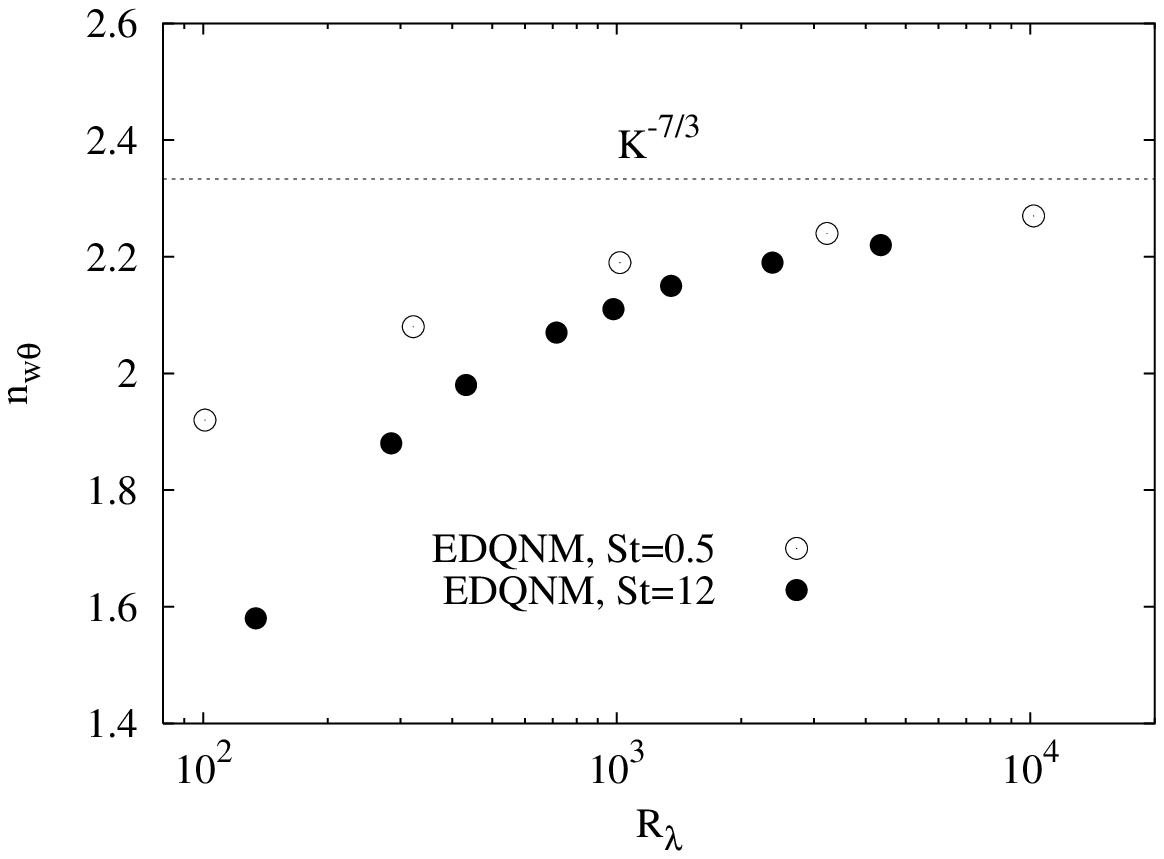}
\includegraphics[scale=0.65,angle=0]{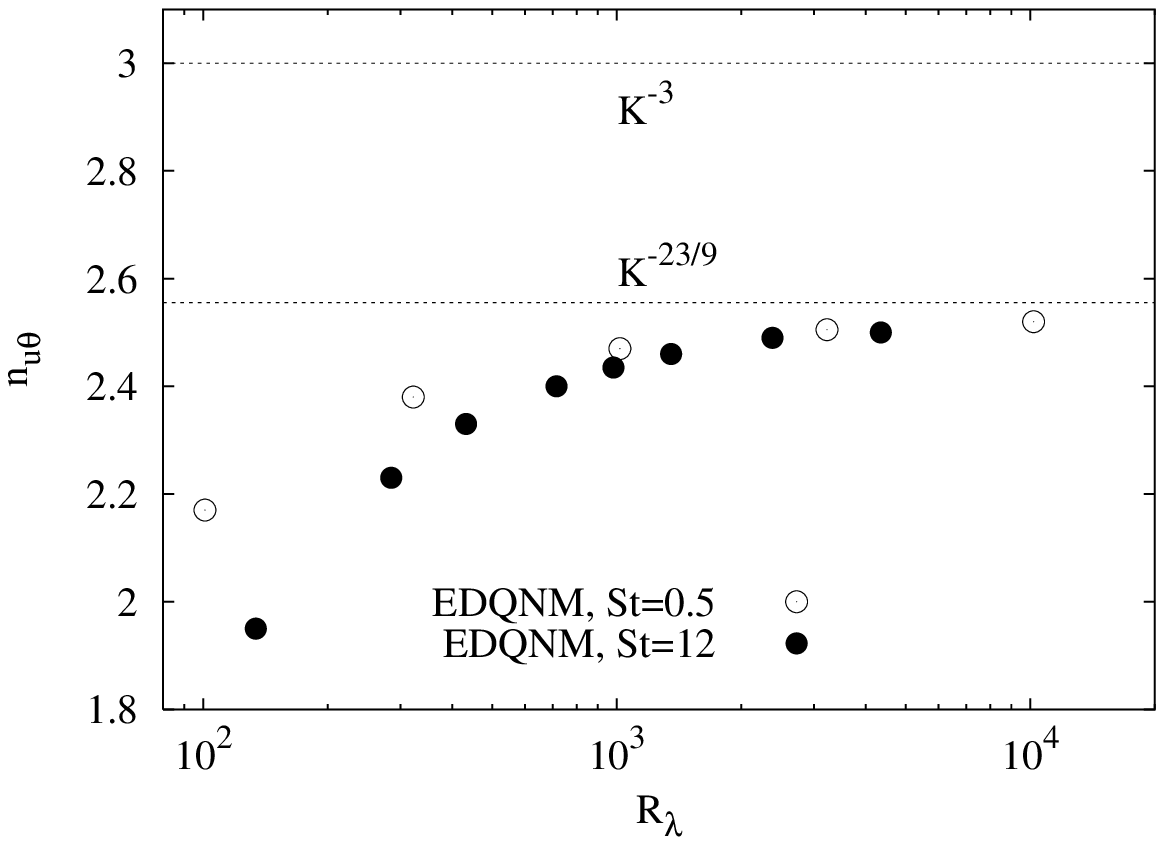}
\caption{Inertial range slopes of the scalar flux spectra as a function of the Reynolds number. Top: cross-stream  scalar flux, the dotted line corresponds to $K^{-7/3}$. Bottom: streamwise scalar flux, the dotted lines correspond to $K^{-23/9}$ and $K^{-3}$.   \label{slopeF1F2}}
\end{figure}

In figure \ref{specF1F2}, the spectra of the streamwise and cross-stream  scalar flux are plotted at $St=0.5$, for six Reynolds numbers in the range $32<R_{\lambda}<10^4$. The streamwise component $F_{u\theta}(K)$, shows a steeper inertial range than $F_{w\theta}(K)$. Figure \ref{slopeF1F2} shows the spectral exponents of these two components of the scalar flux as a function of the Reynolds number. The exponents are determined at two different times, corresponding to $St=0.5$ and $St=12$. It is observed that the difference between the  slopes at different times decreases with increasing Reynolds number. 
{
The spectral exponent tends faster to its asymptotic value for $St=0.5$ than for $St=12$. A possible explanation is the following: the nonlinear triple correlations that are responsible for the creation of an inertial range are initially zero. This yields an initially weak nonlinear transfer, corresponding to a steep spectrum. When the triple correlations are fully developed, the nonlinear transfer becomes stronger, resulting in a less steep spectral distribution.   
}
The cross-stream  scalar flux spectrum tends to the $n_{w\theta}=-7/3$ asymptote for large $R_{\lambda}$, like in an isotropic turbulence with a mean scalar gradient. The asymptotic value is not changed by the presence of shear, as was already anticipated by Lumley \cite{Lumley2}. As is the case in the absence of shear, the asymptote is approached very slowly when increasing the Reynolds number (see the recent paper by the authors \cite{Bos2005}).


\begin{figure}
\includegraphics[scale=0.65,angle=0]{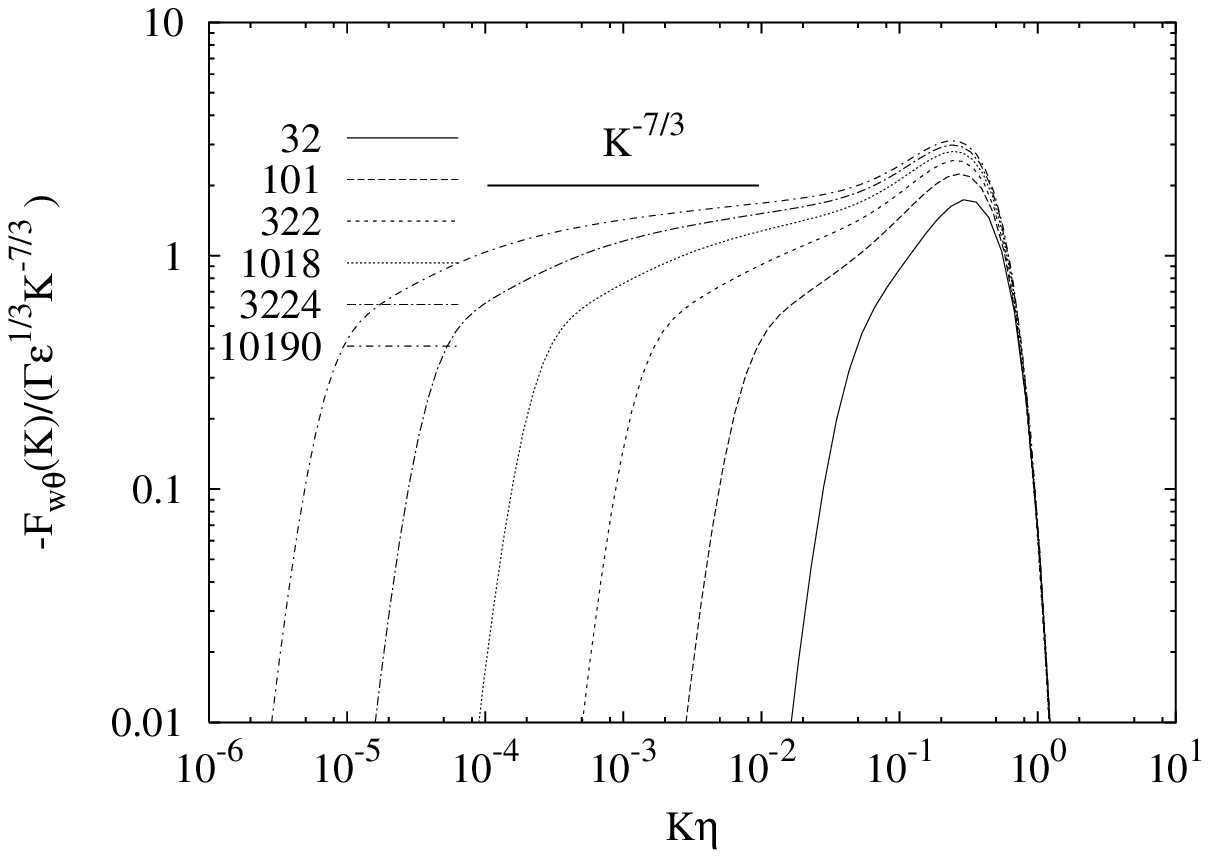}
\includegraphics[scale=0.65,angle=0]{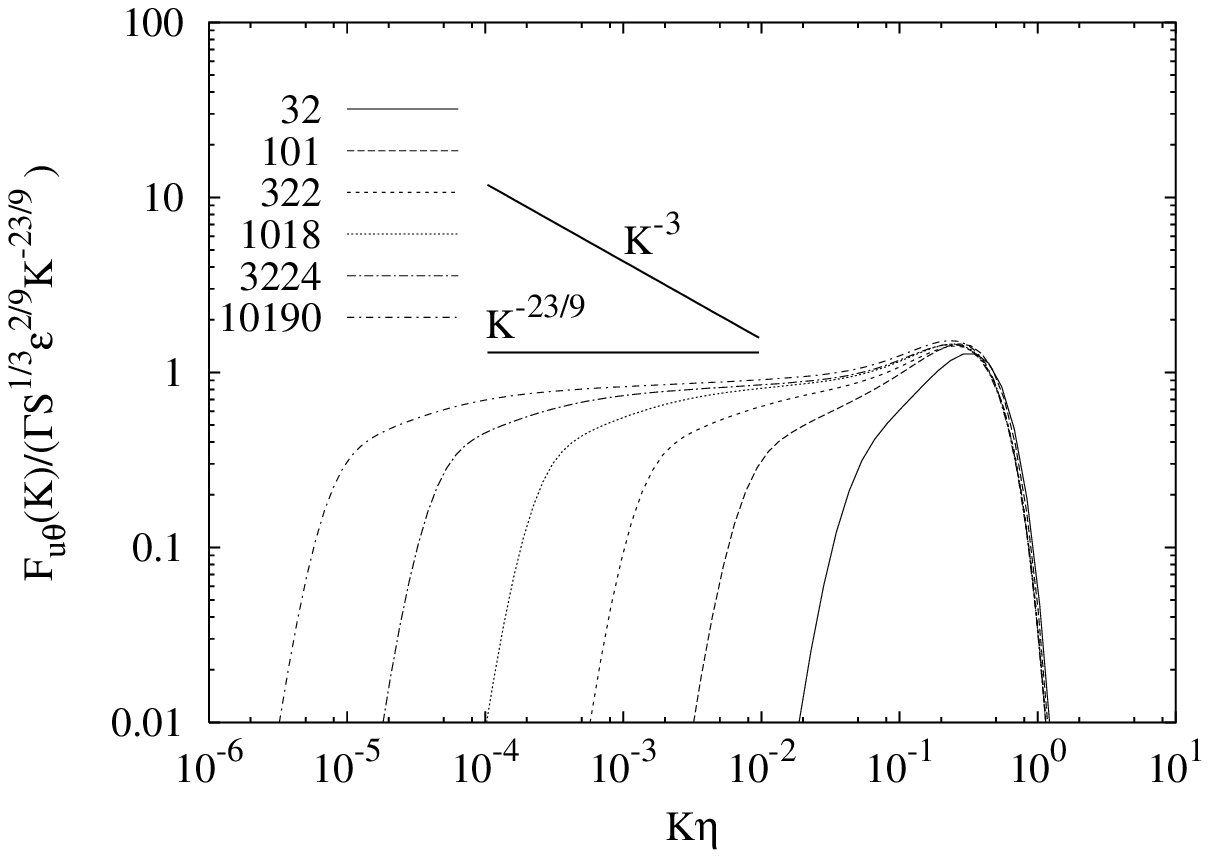}
\caption{Compensated cross-stream (top) and streamwise  (bottom) scalar flux spectra for $32<R_{\lambda}<10^4$ at $St=0.5$. \label{specsF2comp} }
\end{figure}

The streamwise spectrum (fig. \ref{specF1F2}, bottom) is found to be steeper than the cross-stream spectrum. Its exponent, $n_{u\theta}$ (fig. \ref{slopeF1F2}, bottom) tends to a value significantly larger than $7/3$. However, the high Reynolds number asymptote is found to be smaller than the value $n_{u\theta}=3$ proposed by Wyngaard and Cot\'e \cite{Wyngaard}, and $n_{u\theta}={23/9}$ appears a more plausible value. This can also be observed in figure \ref{specsF2comp}, where the spectra compensated by $K^{23/9}$ are plotted. The normalization used in this figure will be clarified in section (\ref{secDima}). 

\begin{figure} 
\begin{center}
\includegraphics[width=0.65\unitlength]{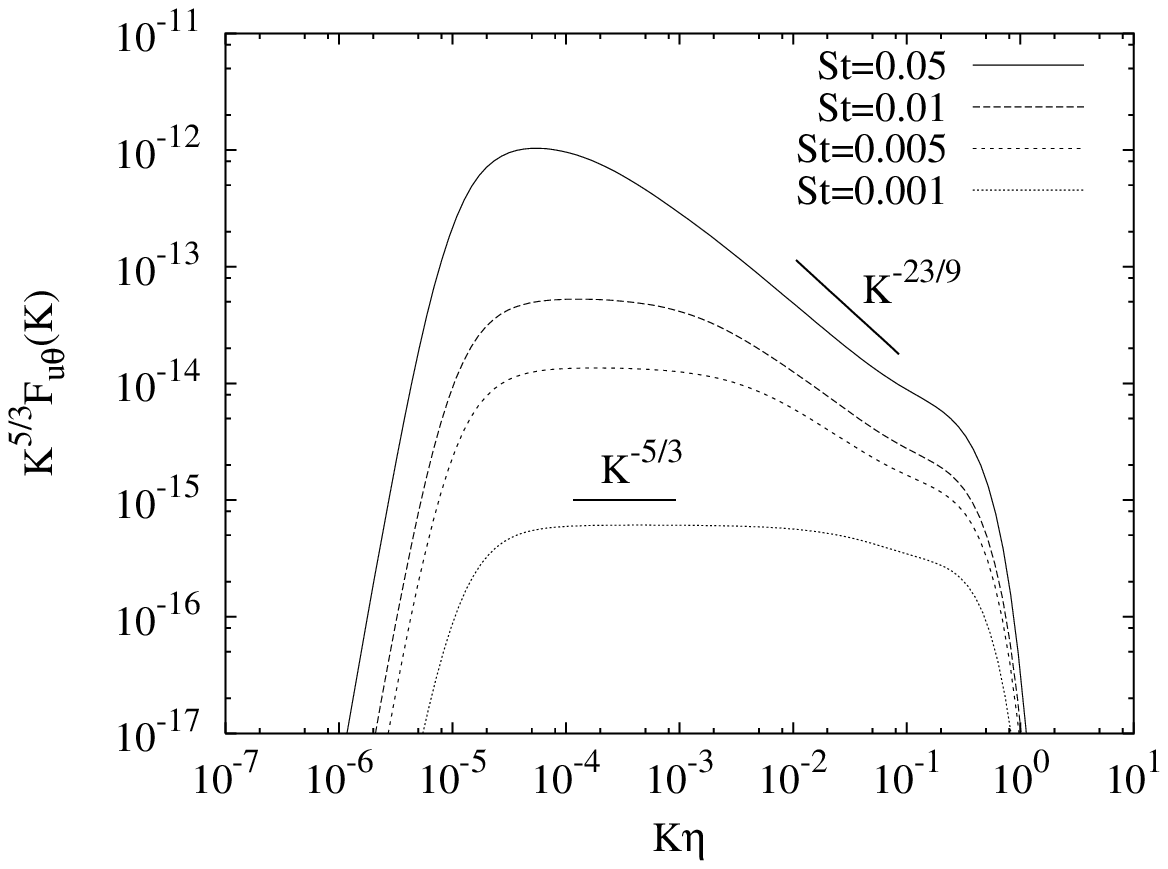}
\end{center}
\caption{The streamwise scalar flux spectrum at four different times in the beginning of the calculation. The spectra are compensated by $K^{5/3}$. \label{figAZSt001}}
\end{figure}

It has to be noticed that the spectra shown and analyzed in this section were obtained at moderate or large nondimensional times, $St=0.5$ and $St=12$. To analyze the behavior at shorter time, the time evolution of the streamwise scalar flux spectrum is plotted in figure \ref{figAZSt001}. It can be observed that at very short times, the spectrum behaves as $K^{-5/3}$, a scaling in agreement with the observations of Antonia and Zhu \cite{AntoniaZhu}, who showed two cases of spectra with a $K^{-5/3}$ scaling for over two decades. The present model therefore can produce a scaling in agreement with the one proposed by these authors. However this scaling is only observed at very short time and does not persist for a longer period and rapidly the spectra become steeper, tending to the asymptotical $K^{-23/9}$ behavior. 
{
We note that, starting from an initial energy spectrum with a $K^{-5/3}$ distribution, the result $n_{u\theta}=-5/3$ can be obtained by purely linear arguments, neglecting all the nonlinear terms.
}

\subsection{Budget of the different terms in the scalar flux spectrum equation \label{Contri}}

\begin{figure}
\includegraphics[scale=0.65,angle=0]{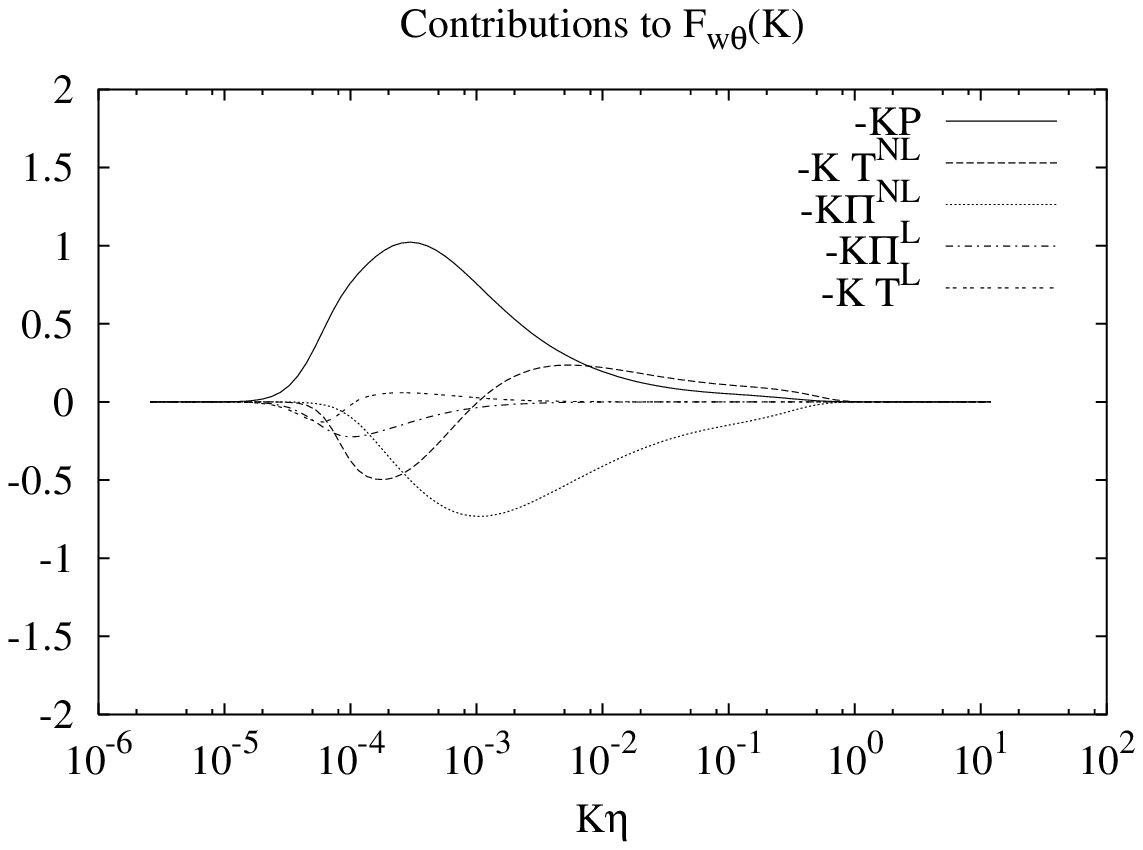}
\includegraphics[scale=0.65,angle=0]{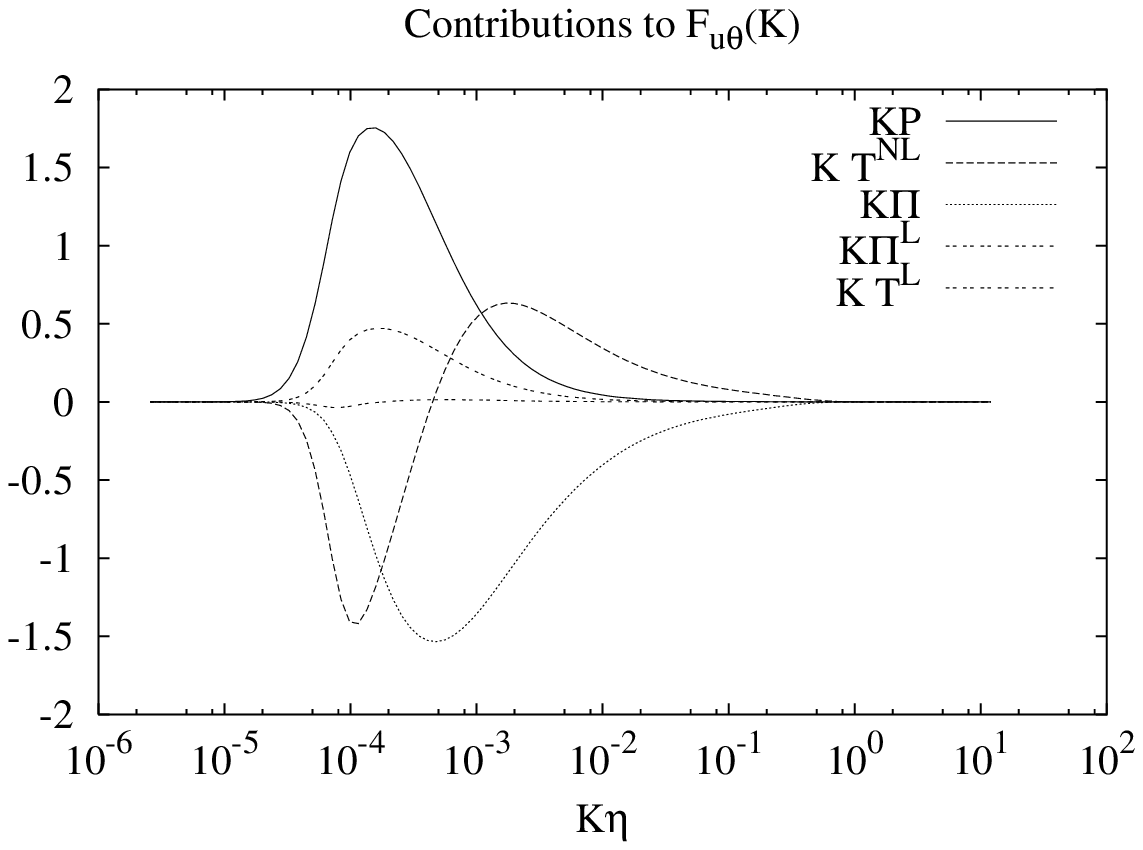}
\caption{Contributions to the equations of the cross-stream and streamwise   scalar flux spectra at $St=12$ and $R_{\lambda}=2384$. \label{termsF1F22384} }
\end{figure}

The different contributions to the evolution equation of the scalar flux spectra,  equation (\ref{eqIntcrF}), are shown in figure 
\ref{termsF1F22384} at $R_\lambda=2384$. To facilitate the analysis and reduce the number of terms, the contributions of the production by the scalar gradient and mean shear were lumped together:
\begin{eqnarray}\label{ProdSdef}
P_i=-\Gamma\varphi_{i3}-S\delta_{i1}{F}_{w\theta}.
\end{eqnarray} 
{
It can be directly deduced from (\ref{ProdSdef}) that there is no explicit contribution from the mean shear to the production term of $F_{w\theta}$.
}
As for the dissipative term, $-(\nu+\alpha) K^2 {F}_{u_i\theta}$, it is negligibly small at this $R_\lambda$ and is therefore not shown.

For simplicity the indices are dropped (all contributions to ${F}_{u\theta}$ should have an index $1$, the contributions to ${F}_{w\theta}$ an index $3$). All the contributions to ${F}_{w\theta}$, which is a negative spectrum, are multiplied by $-1$ to facilitate the comparison with ${F}_{u\theta}$, which is a positive spectrum. 

It can be noted that for both spectra $F_{w\theta}$ and ${F}_{u\theta}$, the main contributions are the production $P$ and the nonlinear transfer and pressure terms, $T^{NL}$ and $\Pi^{NL}$ respectively. The linear transfer $T^L$ and rapid pressure $\Pi^{L}$ terms are relatively small, especially at high wavenumbers. By comparison with Bos \etal \cite{Bos2005}, in which the case of isotropic turbulence with a uniform scalar gradient was analyzed, it is observed that the behavior of $P$, $\Pi^{NL}$ and  $T^{NL}$ is roughly the same. It can be concluded that the presence of the mean shear does not drastically affect the mechanisms involved in the dynamics of $F_{w\theta}$. Compared to the isotropic case, the presence of the mean shear does not directly affect $F_{w\theta}$, as its production remains by the mean scalar gradient only. $F_{w\theta}$ is essentially produced at the large scales and destroyed by the pressure effect. This destruction takes place  locally as well as at smaller scales, since the transfer term introduces a cascade process.  

The streamwise scalar flux spectrum $F_{u\theta}$ does not exist in the isotropic case. It can be observed in figure \ref{termsF1F22384} (bottom) that $F_{u\theta}$ is  essentially generated at large scales by the production term associated with the presence of the mean gradients.  This was already the case for $F_{w\theta}$, but for $F_{u\theta}$ both the mean shear $S$ and the scalar gradient $\Gamma$ contribute to the source term. The transfer term still introduces a cascade effect and the destruction remains associated with the nonlinear pressure. The main difference with the $F_{w\theta}$ budget is that in the case of $F_{u\theta}$, the mean production falls off more rapidly as the wavenumber increases. The two contributions to $P_i$ in equation (\ref{ProdSdef}) indeed behave as $K^{-7/3}$, whereas in the $F_{w\theta}$ case they were found to scale as $K^{-5/3}$. As a result of this more rapid extinction of the production term, it can be observed in figure \ref{termsF1F22384}, that in the inertial range of the spectra, there are only two dominant terms: the nonlinear transfer $T^{NL}$ and the nonlinear pressure term $\Pi^{NL}$.

\section{Dimensional Analysis \label{secDima}}

{
In this section, a dimensional analysis compatible with the results of the spectral closure is proposed. First, the classical dimensional analysis is recalled, and it is stressed that it leads to a spectral scaling that is not compatible with the results of the previous section. Then, based on the observations in section III.c, a new argument is proposed. It leads to a scaling that agrees with the EDQNM results, as well as with existing atmospherical measurements.
}

For the streamwise scalar flux spectrum dimensional analysis based on the quantities $S$, $\epsilon$ and $K$ provides  the following expression for the spectrum:
\begin{equation}\label{eqGSalpha}
F_{u\theta}(K)\sim  \frac{\partial {\overline{\Theta}}}{\partial z} S^{\alpha}\epsilon^{\frac{1-\alpha}{3}} K^{-\frac{7+2\alpha}{3}}.
\end{equation}
{
}
This expression is linear in the scalar gradient, as it has to be to reflect the linearity of the scalar equation. Linearity in $S$ is not mandatory since the Navier-Stokes equations are not linear; if linearity in $S$ is assumed, (\ref{eqGSalpha}) reduces to the Wyngaard and Cot\'e \cite{Wyngaard} formulation:
\begin{equation}
F_{u\theta}(K)\sim \frac{\partial {\overline{\Theta}}}{\partial z} SK^{-3}.
\end{equation}
{
}
This formulation can also be found by assuming an equilibrium between the production term scaling as 
\begin{equation}
P\sim \frac{\partial {\overline{\Theta}}}{\partial z} S \epsilon^{1/3}K^{-7/3} 
\end{equation}
and the pressure destruction term assumed to scale as:
\begin{equation}
-\frac{1}{\tau_{NL}(K)}F_{u\theta}(K),
\end{equation}
in which $\tau_{NL}(K)$ is the nonlinear cascade time
\begin{equation} 
\tau_{NL}(K)\sim \epsilon^{-1/3}K^{-2/3}.
\end{equation}
The analysis of the different contributions in figure \ref{termsF1F22384} having shown that the nonlinear terms are dominant in the inertial range and that the local production term plays a negligible role, an alternative reasoning, based on a wavenumber dependent flux (cf. Bos \etal \cite{Bos}) could be defendable. For this reason it is proposed  to introduce a tensorial extension of (\ref{lumleyscaling}):
\begin{equation}\label{MY12}
F_{u_i\theta}(K)\sim \frac{\partial {\overline{\Theta}}}{\partial x_j} \Delta_{ij}(K)K^{-7/3},
\end{equation}
{
in which $\Delta_{ij}(K)$ is a quantity that reflects the anisotropic character of the nonlinear cascade and has the dimension of $\epsilon^{1/3}$. We define $\Delta$ such that each of its components is the cube root of the corresponding component of $\epsilon_{ij}$: $\Delta_{ij}\rightarrow\epsilon_{ij}^{1/3}$,
}
in which $\epsilon_{ij}$ stands for a tensorial spectral flux, which is not necessarily conserved in the cascade: $\epsilon_{ij}=\epsilon_{ij}(K)$. It can be assumed that $\epsilon_{ij}(K)$ is a spectral anisotropic flux in wavenumber space, whose anisotropy will be proportional to the anisotropy of the spectral tensor $\varphi_{ij}(K)$:  
\begin{equation}\label{eq26}
\epsilon_{ij}(K)=\frac{\frac{3}{2}\varphi_{ij}(K)}{E(K)}\epsilon, 
\end{equation}
in which $\epsilon$ represents the total energy flux and $E(K)$ is the turbulent kinetic energy spectrum. Using Kolmogorov scaling for the energy spectrum and Lumley's expression\cite{Lumley},
\begin{equation}\label{uwspec}
\varphi_{uw}(K)\sim S \epsilon^{1/3}K^{-7/3}.
\end{equation}
 for the Reynolds-stress spectrum, expression (\ref{eq26}) yields in the inertial range:
\begin{eqnarray}
\epsilon_{ww}(K)=\epsilon\textrm{,}\nonumber \\
\epsilon_{uw}(K)\sim S\epsilon^{2/3}K^{-2/3}
\end{eqnarray}
which yields the Lumley scaling (\ref{lumleyscaling}) for the cross-stream scalar flux and for  $F_{u\theta}(K)$:
\begin{equation}\label{239}
F_{u\theta}(K)\sim  \frac{\partial {\overline{\Theta}}}{\partial z}S^{1/3}\epsilon^{2/9} K^{-23/9}.
\end{equation}
This scaling is in full agreement with the results presented in this paper. 
Noticing that $23/9 \approx 2.55$, it is also in agreement with atmospheric measurements leading to $n_{u\theta}\approx 2.5$. We insist here however that the presence of the parameter $\alpha$ in (\ref{eqGSalpha}) allows for a wide range of possible scalings, so that further research is needed to confirm or disqualify the scaling (\ref{239}).
{
An interesting feature of expression (\ref{239}) is that it is not linear dependent on the mean velocity gradient. The implications of this for one-point models of the scalar flux deserve more attention.

}

\section{Conclusion}

Two point closure theory was applied to study the scalar flux in homogeneously sheared turbulence with a cross-stream scalar gradient. The equation for the spectrum $F_{u_i\theta}(K)$ was closed by EDQNM theory and tensor invariant modeling. The resulting model was used to study the Reynolds number dependency of the spectra associated to the cross-stream and streamwise scalar flux. The asymptotic value of the spectral exponent of the cross-stream scalar flux spectrum is shown to be $-7/3$, similar to the case without shear. The spectral exponent of the streamwise scalar flux spectrum is shown to approach a value close to a $-23/9$ scaling, disagreeing with classical dimensional analysis which predicts a $-3$ scaling. The $-23/9$ scaling  appears to be in better agreement with atmospheric observations.

A new dimensional argument was then proposed, depending on an anisotropic spectral energy flux. It leads to an expression in agreement with both the $K^{-23/9}$ behavior of the cross-stream component and a $K^{-7/3}$ dependency for the streamwise spectrum.

It is sometimes considered that two-point closure theories can only yield trivial results, entirely predictable by dimensional arguments. The possible existence of the $K^{-23/9}$ scaling illustrates the opposite. Indeed, this scaling can be obtained by dimensional analysis as shown in section \ref{secDima}, but a $K^{-3}$ scaling can also be obtained by dimensional analysis. The authors hope that the present two-point closure results might inspire future work, which prove or disprove the scaling (\ref{239}).

\begin{acknowledgements}
Fruitful discussions with Claude Cambon, Jean-No\"el Gence, Fabien Godeferd and Yohann Duguet are gratefully acknowledged. Hatem Touil is acknowledged for providing the numerical code for anisotropic sheared turbulence used in this work, as well as for his help in analyzing the results. 
{
The referees are acknowledged for their constructive comments.}
\end{acknowledgements}



\end{document}